\documentclass[sigplan,10pt]{acmart}

\usepackage{float}
\usepackage{multirow}
\usepackage{comment}
\usepackage{lineno}
\usepackage{xcolor,colortbl}
\usepackage{tabularx}
\usepackage{multirow}
\usepackage{color}
\usepackage{listings}
\usepackage{lstautogobble}
\usepackage{xspace}
\usepackage{amsmath}
\usepackage{hyperref}
\usepackage{soul}


\usepackage{subcaption}
\usepackage{caption}

\definecolor{light}{rgb}{0.5, 0.5, 0.5}
\def\light#1{{\color{light}#1}}

\newcommand{\name}{StarPlat\xspace} 

\newcommand{\keywordSP}[1]{{\color{blue}{#1}}}

\newcommand{\bcc}{Betweenness Centrality\xspace} 
\newcommand{\prr}{PageRank\xspace} 
\newcommand{\ssspp}{Single-Source Shortest Paths\xspace} 
\newcommand{\tcc}{Triangle Counting\xspace}

\newcommand{\MLE}{\color{red}{OOM}}
\newcommand{\OOT}{\color{red}{OOT}}

\newcommand{\rtodo}[1]{{\color{blue}{rajz: #1}.}}

\newcommand{\mypara}[1]{\vspace{1mm}\noindent\textbf{#1.} }
\newcommand{\REM}[1]{}

\begin{document}

\title{Code Generation for a Variety of  Accelerators for a Graph DSL}

\author{Ashwina Kumar}
\email{cs20d016@cse.iitm.ac.in}
\orcid{1234-5678-9012}
\affiliation{%
  \institution{IIT Madras}
  \country{India}
}

\author{M. VENKATA KRISHNA}
\email{mvkgta26@gmail.com}
\affiliation{%
  \institution{PSG Tech}
  \country{India}
}
\author{Prasanna Bartakke}
\email{prasanna.bartakke@gmail.com}
\orcid{0009-0005-8129-397X}
\affiliation{%
  \institution{IIT Madras}
  \country{India}
}
\author{Rahul Kumar}
\email{rahul7870144@gmail.com}
\affiliation{%
  \institution{IIT Madras}
  \country{India}
}
\author{Rajesh Pandian M}
\email{mrprajesh@cse.iitm.ac.in}
\orcid{0000-0003-4702-4678}
\affiliation{%
  \institution{IIT Madras}
  \country{India}
}

\author{Nibedita Behera }
\email{cs20s023@cse.iitm.ac.in}
\orcid{0000-0002-1563-8686}
\affiliation{%
  \institution{IIT Madras}
  \country{India}
}

\author{Rupesh Nasre}
\email{rupesh@cse.iitm.ac.in}
\orcid{0000-0001-7490-625X}
\affiliation{%
  \institution{IIT Madras}
  \country{India}
}

\renewcommand{\shortauthors}{Kumar, et al.}




\begin{abstract}
Sparse graphs are ubiquitous in real and virtual worlds. With the phenomenal growth in semi-structured and unstructured data, sizes of the underlying graphs have witnessed a rapid growth over the years. Analyzing such large structures necessitates parallel processing, which is challenged by the intrinsic irregularity of sparse computation, memory access, and communication. It would be ideal if programmers and domain-experts get to focus only on the sequential computation and a compiler takes care of auto-generating the parallel code. On the other side, there is a variety in the number of target hardware devices, and achieving optimal performance often demands coding in specific languages or frameworks. Our goal in this work is to focus on a graph DSL which allows the domain-experts to write almost-sequential code, and generate parallel code for different accelerators from the same algorithmic specification. In particular, we illustrate code generation from the StarPlat graph DSL for NVIDIA, AMD, and Intel GPUs using CUDA, OpenCL, SYCL, and OpenACC programming languages. Using a suite of ten large graphs and four popular algorithms, we present the efficacy of StarPlat's versatile code generator.

\end{abstract}


\lstdefinelanguage{NEAR}
{
  morekeywords= [1]{ COPY,
  ENTRYPOINT, function, fixedPoint, filter, Min, forall, iterateInBFS, iterateInReverse, submit, parallel_for, wait, pragma, acc, data, copyin, copy, copyout, parallel, loop, __global__, __kernel, clEnqueueNDRangeKernel, clWaitForEvents, atomicMin, reduction, atomic_ref, atomic, write, fetch_min},
  morekeywords = [2]{propNode, Graph, node, edge, int, float, bool, in, until, for, if, from, do, while, void, unsigned},
  morekeywords = [3]{attachNodeProperty, nodes, neighbors, get_edge},
  morekeywords = [4]{True, False},
  morecomment=[l]{\//},
  morestring=[b]",
}

\renewcommand{\lstlistingname}{Fig}
\lstset{escapeinside={/*@}{*/}}
\lstdefinestyle{nolinenum}{
numbers=none
}
\lstdefinestyle{mystyle}{
  xleftmargin=.1\textwidth, 
  numbers=left, 
  numbersep=5pt,                  
  basicstyle=\small,
  backgroundcolor=\color{white},  
  xleftmargin=.01\textwidth,
  showspaces=false,
  rulecolor=\color{black},
  stringstyle=\color{green},
  numberstyle=\scriptsize\color{gray},
  commentstyle=\color{gray},
  identifierstyle=\color{black},
  keywordstyle = [1]\color{purple},
  keywordstyle=[2]\color{blue},
  keywordstyle=[3]\color{black},
  keywordstyle=[4]\color{green},
  keywordstyle=\color{purple}\bfseries,
  showstringspaces=false,        
  showtabs=false,                
  tabsize=1,                      
  captionpos=b,                   
  breaklines=true,                
  breakatwhitespace=true,         
  title=\lstname, 
  float=tp,
  floatplacement=tbp,
}
\lstdefinestyle{cstyle}{
 xleftmargin=.1\textwidth, 
  numbers=left,
  numbersep=8pt,                  
  backgroundcolor=\color{white},  
  showspaces=false,
  rulecolor=\color{black},
  stringstyle=\color{green},
  numberstyle=\scriptsize\color{gray},
  commentstyle=\color{gray},
  identifierstyle=\color{black},
  keywordstyle=\color{purple},
  showstringspaces=false,        
   showtabs=false,                
  tabsize=1,                      
  captionpos=b,                   
  breaklines=true,                
  breakatwhitespace=true,         
  title=\lstname 
}

\maketitle
\section{Introduction} \label{sec intro}
Graphs naturally model several real-world phenomena such as social networks, road networks and biological systems. Computer systems today need to deal with huge graphs: Facebook has 2.9 billion monthly active users, US has a rail network of 260,000 km with a transport of 10.3 billion passenger-km, while our brain network has 86 billion neurons. To deal with such a scale, it is natural to employ farms of parallel machines housing powerful accelerators. We continue to witness various architecture designers and vendors proposing new hardware to enable us process huge data fast.

To accelerate our algorithms and applications optimally on a certain hardware, currently, the domain expert (such as a biologist) needs to program in a certain language or a framework geared towards the hardware. For instance, NVIDIA GPUs are tied closely with CUDA, Intel GPUs expect SYCL, while AMD GPUs need Hip for extracting the best performance. This is clealy not ideal. Therefore, we have been moving towards common computing languages such as OpenCL and OpenACC, which are supposed to be platform portable. Unfortunately, their support is currently limited, and sometimes driven by vendor competition.

In this work, we focus on graph algorithms and accelerators, and help domain-experts generate parallel code for multiple accelerators from the same algorithmic specification. In this endeavour, we employ a recently proposed domain-specific language for graph algorithms named StarPlat~\cite{starplat} and augment its compiler to support multiple accelerator backends: CUDA, SYCL, OpenCL and OpenACC. 

This paper makes the following contributions.
\begin{itemize}
\item A code generator which builds upon the intermediate representation of StarPlat to support NVIDIA, Intel, and AMD GPUs. This relieves the domain-expert from learning new languages to write parallel code for graph algorithms. 
\item Custom processing per accelerator for optimized computation. This allows StarPlat to, for instance, optimize data-clause around loops in OpenACC, while optimize reduction for the SYCL backend.
\item An extensive experimental evaluation of the generated codes against library-based Gunrock and manually-optimized LonestarGPU on ten large graphs and four popular graph algorithms (betweenness centrality, page rank, single-source shortest paths, and triangle counting), illustrating StarPlat's ability to be competitive with hand-crafted codes.
\end{itemize}


The rest of the paper is organized as follows: Section~\ref{sec background} presents the language specification and various accelerator backends along with an example DSL program. Section~\ref{sec code gen} describes the code-generation scheme followed for the translation of the DSL code for each backend accelerator.
Section~\ref{sec optimizations} provides an overview of the backend-specific optimizations employed for efficient code generation. Section~\ref{sec exp evaluation} presents the experimental evaluation of the generated code for each backend. Section~\ref{sec relatedwork} discusses the related work for graph analytics. We summarise our experience and conclude in Section~\ref{sec conclusion}.
\section{Background} \label{sec background}
Figure~\ref{bc-dsl-sample} presents the algorithmic specification for computing betweenness centrality (BC) in the StarPlat DSL. It employs Brandes algorithm \cite{brandes} resembling unweighted all pairs shortest paths (APSP). To be practically tractable, literature usually runs a few iterations of APSP, which can be specified in StarPlat as \textsf{sourceSet} to the \textsf{ComputeBC} function. For each such source (Line~\ref{Compute_BC-stat-4}), the algorithm performs a forward and a backward pass accumulating \textsf{sigma} (number of shortest paths) in the forward pass and updating delta values (fraction of the shortest paths) in the backward pass.

\begin{lstlisting}[language=NEAR, style=mystyle, label=bc-dsl-sample, caption = BC specification in \name %~\rtodo{ RN, Is running Fig+Code numbers  okay? Codes named as Figs is ok?}
%, xleftmargin=.01\textwidth
]
function /*@\textbf{ComputeBC}*/ (Graph g, propNode <float> BC, SetN<g> sourceSet) {
  g.attachNodeProperty (BC = 0); /*@\label{Compute_BC-stat-3}*/ 
 
  for (src in sourceSet) { /*@\label{Compute_BC-stat-4}*/                  
    propNode<float> sigma; /*@\label{Compute_BC-stat-5}*/ 
    propNode<float> delta; /*@\label{Compute_BC-stat-6}*/ 
    g.attachNodeProperty(delta = 0); /*@\label{Compute_BC-stat-7}*/ 
    g.attachNodeProperty(sigma = 0); /*@\label{Compute_BC-stat-8}*/ 
    src.sigma = 1; /*@\label{Compute_BC-stat-9}*/ 
         
    iterateInBFS(v in g.nodes() from src){ /*@\label{Compute_BC-stat-10}*/
      for(w in g.neighbors(v)) { /*@\label{Compute_BC-stat-11}*/
        v.sigma = v.sigma + w.sigma; /*@\label{Compute_BC-stat-12}*/ 
      }  /*@\label{Compute_BC-stat-13}*/ 
    }  /*@\label{Compute_BC-stat-14}*/ 
    iterateInReverse(v != src) { /*@\label{Compute_BC-stat-15}*/ 
      for(w in g.neighbors(v)) { /*@\label{Compute_BC-stat-16}*/
     	  v.delta = v.delta + (v.sigma / w.sigma) * (1 + w.delta); /*@\label{Compute_BC-stat-17}*/ 
      }
      v.BC = v.BC + v.delta; /*@\label{Compute_BC-stat-18}*/ 
} } } /*@\label{Compute_BC-stat-19}*/ 
\end{lstlisting}

The function takes three parameters: the underlying graph, BC values to be updated, and the sources from where the shortest path computation is to be initiated. Line~\ref{Compute_BC-stat-3} uses \textsf{attachNodeProperty} which initializes the BC value of each node to 0. The \textsf{for} loop at Line~\ref{Compute_BC-stat-4} iterates through all the sources in \textsf{sourceSet}.
Lines~\ref{Compute_BC-stat-5} and \ref{Compute_BC-stat-6} define new node attributes, which are initialized in
Lines~\ref{Compute_BC-stat-7} and \ref{Compute_BC-stat-8}. 
Lines~\ref{Compute_BC-stat-10} to \ref{Compute_BC-stat-14} represent forward BFS using the \textsf{iterateInBFS} construct, while  Lines~\ref{Compute_BC-stat-15} to \ref{Compute_BC-stat-19} represent backward BFS using \textsf{iterateInReverse} (which must be preceded by \textsf{iterateInBFS}).


\subsection{StarPlat Language Constructs}
We briefly discuss StarPlat's language constructs~\cite{starplat}.

\mypara{Data Types}
\name supports primitive data types: \texttt{int}, \texttt{bool}, \texttt{long}, \texttt{float}, and \texttt{double}.
It also supports Graph, node, edge, node attribute, edge attribute, etc. as first-class types.

\mypara{Parallelization and Iteration Schemes}
\texttt{forall} is an aggregate construct that can process a set of elements in parallel. Its sequential counterpart is a simple \light{for} statement. Currently, \name supports vertex-based processing.

\mypara{Reductions}
While reduction is often a fundamental building block of parallel languages to enable synchronization, supporting reduction as an extra language construct does not directly align with \name's language design. Therefore, the combined operators such as += are used to specify reduction, preserving the abstraction. 
The reduction operators supported by \name are tabulated in Table~\ref{tab:reduction}.

\begin{table}
\centering
\captionsetup{justification=centering}
\begin{center}
\begin{tabular}{ |c|c| } 
 \hline
 {\textbf{Operator}} & {\textbf{Reduction Type}}\\
 \hline
 {\texttt{+=}} & {Sum} \\
 \hline 
 {\texttt{*=}} & {Product} \\ 
 \hline 
 {\texttt{++}} & {Count} \\
 \hline 
  {\texttt{\&\&=}} & {All}\\
 \hline 
 {\texttt{||=}} & {Any} \\
 \hline 
 
\end{tabular}
\caption{Reduction operators in \name}
\label{tab:reduction}
\end{center}
\end{table}



\mypara{FixedPoint and Min/Max Constructs}
Several solutions to graph algorithms are iterative and converge based on conditions on the node attributes.
\name provides a fixedPoint construct to specify this succinctly. Its syntax involves a boolean variable and a boolean expression on node-properties forming the convergence condition, as shown below.

\noindent\texttt{\keywordSP{fixedPoint until} (var: convergence expr) \{...\}}


\name provides constructs \texttt{Min} and \texttt{Max} which perform multiple assignments atomically based on a comparison criterion. This can be useful in update-based algorithms like SSSP, where an update on node properties is carried out on a  desired condition while taking care of potential dataraces.


\name also provides aggregate functions \texttt{minWt} and \texttt{maxWt} to find the minimum and the maximum edge weights.

\subsection{Accelerator Programming Backends}
We briefly discuss various accelerator backends supported. 

\mypara{CUDA}
CUDA is a programming model created by NVIDIA for programming its graphics processing units (GPUs). It is used to speed up computationally demanding activities in a variety of industries, including machine learning, scientific computing, and video game creation. Due to the high computing needs of training deep neural networks, CUDA has grown to be a prominent platform for accelerating deep learning algorithms.

\mypara{SYCL}
SYCL adds data parallelism to C++. It is a standard language for defining data-parallel computations that can be run on many hardware platforms. Several hardware manufacturers, including Intel, AMD, Xilinx, and NVIDIA, support SYCL. SYCL provides a higher-level abstraction that enables programmers to use modern C++ features such as templates and lambda expressions while still writing code in the traditional C++ language.
Moreover, SYCL offers a standard library of algebraic and mathematical functions that can be applied to high-performance computing. 

\mypara{OpenCL}
Applications that need parallel processing can be accelerated with the help of OpenCL, which offers a consistent programming model.
Without needing to write a separate code for every device, it enables developers to create code that can run on a variety of hardware. OpenCL is similar in spirit to StarPlat, but has a limited types of devices supported (e.g., OpenCL cannot parallelize directly in a distributed setting) and cannot take advantage of device-specific features (e.g, warp level intrinsics in NVIDIA GPUs).

\mypara{OpenACC}
Applications can be accelerated with OpenACC on a variety of heterogeneous computing platforms, including GPUs and CPUs.
The goal of OpenACC is to offer a high-level parallelization solution that is both portable and performance portable, i.e., the code may run on several hardware platforms and yet deliver satisfactory performance.
A programmer can indicate which portions of the code should be parallelized and how they should be parallelized using a set of OpenMP-like directives that are provided by the OpenACC standard.
The standard also includes a set of APIs for data management and device management, which allow the programmer to regulate and optimize data transfer between the host and the accelerator (GPU or FPGA).
\section{StarPlat Code Generator} \label{sec code gen}
We augment StarPlat's code generator to support efficient code generation for CUDA, OpenCL, SYCL, and OpenACC.
In this section, we discuss key design decisions and the challenges faced during code generation. 

\subsection{Graph Representation and Storage}
Graphs enjoy a variety of representations. In the context of StarPlat, the following requirements drove our decision towards a suitable graph representation.
\begin{itemize}
    \item should work across all the accelerators, and preferably on CPU too
    \item should work well with vertex-centric algorithms, common in graph processing
    \item should be compact (minimizing the extra memory) 
    \item should be fast accessible
\end{itemize}
The compressed sparse row (CSR) storage format satisfied our requirements well. Since it uses offset-based arrays, the same memory representation works across all the accelerators as well as the CPU. Adjacency matrix was unable to scale to large graphs, adjacency list involved pointer chasing which reduced efficiency, edge-list was compact and fast, but unsuitable for commonly used vertex-based processing.\footnote{We advocate edge-list based representation for edge-centric graph processing (e.g., Kruskal's MST) across accelerators.}  

\subsection{Neighborhood Iteration}
We discuss accelerator-specific neighborhood traversal below.
Figure~\ref{nbritr-cuda} illustrates it for NVIDIA GPUs. A challenge in CUDA is that the kernel launch and the kernel function are separate. This demands a split-code generation, unlike other backends. This is non-trivial because the StarPlat source code uses simply a \texttt{forall} loop. The CUDA code generator needs to identify the variables used in the kernel, transfer those to the device (\texttt{cudaMemcpy}) and pass those as parameters to the kernel.
The kernel uses the CSR representation (offset array \texttt{gpu\_OA}) to traverse through the neighbors. 

\begin{lstlisting}[language=NEAR, style=mystyle, label= nbritr-cuda,  caption = {CUDA code  generated for neighborhood iteration}]
__global__ void /*@\textbf{computeSSSP} */(...) {
  unsigned id = blockIdx.x * blockDim.x + threadIdx.x;
  ...
  for (int ee = gpu_OA[id]; ee < gpu_OA[id+1]; ee++) {
    int nbr = g.gpu_edgeList[ee];
    ...
}...}...
/*@\textbf{computeSSSP}*/<<<nblocks,blocksize>>>(...);
cudaDeviceSynchronize();
\end{lstlisting}

Figure~\ref{nbritr-openacc} shows neighbor iteration in OpenACC using \texttt{\#pragma acc} annotations. The code generator does not need to create a separate kernel, but faces a non-triviality similar to that in CUDA, since the variables used within the loop need to be \textit{promoted} up to a \texttt{data copyin} clause, to avoid repeated data transfer (e.g., edge weights, vertex attributes). 
        
\begin{lstlisting}[language=NEAR, style=mystyle, label= nbritr-openacc,  caption = {OpenACC code generated for neighborhood iteration}]
#pragma acc data copyin(g)
{
  #pragma acc data copyin( g.edgeList[0:g.num_edges()], g.indexofNodes[:g.num_nodes()+1], weight[0: g.num_edges()], modified[0: g.num_nodes()], modified_nxt[0:g.num_nodes()]) copy(dist[0:g.num_nodes()])
  { ... 
    #pragma acc parallel loop
    for(int v=0; v<g.num_nodes(); v++) {
        for (int ee = g.indexofNodes[v]; ee<g.indexofNodes[v+1]; ee++) {
          int nbr = g.edgeList[ee];
          ...
    }...} ...
} }

\end{lstlisting}

Figure~\ref{nbritr-sycl} shows neighbor iteration
in SYCL expressed as a data-parallel kernel using the \texttt{parallel\_for} function. The first argument, NUM\_THREADS, is the number of items to launch in parallel. The second argument is the kernel function to be executed by each work item, which processes up to $|$V$|$/NUM\_THREADS number of nodes of the graph. 
\lstinputlisting[language=NEAR, style=mystyle, label= nbritr-sycl,  caption = {SYCL code  generated for neighborhood iteration}]
{code/sycl/nbritr.txt}

Figure~\ref{nbritr-opencl} shows neighbor iteration in OpenCL, which, similar to CUDA, has a separate kernel and host processing, with a different syntax.
As observed, while the parallelism concepts remain the same, the syntax and the placement of constructs change significantly across the backends.

\begin{lstlisting}[language=NEAR, style=mystyle, label= nbritr-opencl,  caption = {OpenCL code  generated for neighborhood iteration}]
__kernel void /*@\textbf{computeSSSP}*/(...){
    unsigned id = get_global_id(0);
    ...
    for(int ee = gpu_OA[id]; ee < gpu_OA[id+1]; ee++){
        int nbr = gpu_edgeList[ee];
        ...
    } ...
} ...
// from host
status = clEnqueueNDRangeKernel(command_queue, computeSSSP, 1, NULL, global_work_size, local_work_size, 0, NULL, &event);
clWaitForEvents(1, &event);
\end{lstlisting}

\subsection{Reductions}
We discuss accelerator-specific code generation for reduction below.
Due to restrictions on the number of resident thread-blocks that might block the kernel, the situation in CUDA becomes challenging. Utilizing a reduction from the host as a distinct kernel (for example, using \texttt{thrust::reduce}) is one way to combat this. However, doing so necessitates ending the kernel, leaving the host, calling reduce, returning to the host, and then starting a new kernel to complete the remaining tasks in the \texttt{forall}  loop. In addition to complicating code production, this makes processing as a whole, ineffective.
So, we rely on CUDA atomics to generate the functionally equivalent code (e.g., using \texttt{atomicAdd} as shown in Figure~\ref{reduction-cuda} for operator += as used in triangle counting). 

\begin{lstlisting}[language=NEAR, style=mystyle, label=reduction-cuda, caption = CUDA code generated for reduction / \texttt{Min} construct]
int nbr = gpu_edgeList[edge];
int e = edge;
int dist_new = gpu_dist[v] + gpu_weight[e];

if (gpu_dist[nbr] > dist_new) {
    atomicMin(&gpu_dist[nbr], dist_new);
    gpu_modified_next[nbr] = true;
    gpu_finished[0] = false;
}
\end{lstlisting}

In the case of OpenACC, 
the loops with a reduction operation are marked with a \texttt{reduction pragma}. The reduction clause in OpenACC is helpful in parallelising in such situations to avoid data races. 
An example of OpenACC's reduction clause is shown in Figure~\ref{reduction-openacc}. 

\begin{lstlisting}[language=NEAR, style=mystyle, label= reduction-openacc,  caption = {Generated OpenACC reduction code in PageRank}]
#pragma acc parallel loop reduction(+:diff)
for (int v = 0; v < g.num_nodes(); v++) {
  float sum = 0.0;
  for (int ee = g.rev_indexofNodes[v]; ee < g.rev_indexofNodes[v+1]; ee++) {
    int nbr = g.srcList[ee] ;
    sum = sum + pageRank[nbr] / (g.indexofNodes[nbr+1] - g.indexofNodes[nbr]);
  }
  float val = (1 - delta) / num_nodes + delta * sum;
  diff = diff + val - pageRank[v];
  pageRank_nxt[v] = val;
}
\end{lstlisting}

Figure~\ref{redn-sycl} shows the usage of reduction in SYCL for Triangle Counting algorithm. It relies on the \texttt{atomic\_ref} type for achieving data-race-free update to the global count. 

\begin{lstlisting}[language=NEAR, style=mystyle, label= redn-sycl,  caption = {SYCL code  for reduction in Triangle Counting}]
Q.submit([&](handler &h){ 
  h.parallel_for(NUM_THREADS, [=](id<1> v){
    for (; v < V; v += NUM_THREADS){
      for (int ee = g.gpu_indexOfNodes[v]; ee < g.gpu_indexOfNodes[v+1]; ee++) {
        int u = g.gpu_edgeList[ee];
        if (u < v){ 
          ...
          int w = ...
          if (v < w){
            if(findNeighborSorted(u,w,...)){ 
        atomic_ref<long, memory_order::relaxed, memory_scope::device, access::address_space::global_space> atomic_data(dev_triangle_count[0]);
              atomic_data += 1;
  }  }  }  }  } }
); }).wait();
\end{lstlisting}

Similar to other backends (CUDA and SYCL), OpenCL also uses atomics to implement reduction. Since OpenCL supports atomics only for \texttt{int} and \texttt{long} types, we have simulated those for \texttt{float} and \texttt{double} using \texttt{atomic\_cmpxchg}.


\subsection{BFS Traversal}
The \texttt{iterateInBFS} construct in StarPlat implements Breadth-First Search (BFS) to traverse a graph in parallel.
Similar to the case of neighborhood iteration, the CUDA backend expects the code generated for the \texttt{iterateInBFS} construct to be separate for the host and the device.
The level-wise BFS kernel on the GPU is called internally by the outer \texttt{do-while}  loop, which operates on the host (Line~\ref{line:bfsdowhile}). The flag (\texttt{finished}) is copied across devices (Lines~\ref{line:bfsh2d} and \ref{line:bfsd2h}), passed as a parameter to the kernel, and updated in the kernel whenever there is a change in a vertex's level because the loop is on the host.

\begin{lstlisting}[language=NEAR, style=mystyle, label=bfs-cuda, caption = CUDA code for the \texttt{iterateInBFS} construct]
do {          /*@\label{line:bfsdowhile}*/
  H2D(...); /*@\label{line:bfsh2d}*/
  BFS<<<...>>>(...)
  cudaDeviceSynchronize();
  ++hops_from_source;

  D2H(...); /*@\label{line:bfsd2h}*/
  ...
} while (!finished);

__global__ void BFS(...) { /*@\label{line:bfskernel}*/
  ... 
  if (d_level[u] == *d_hops_from_source) {
    ... 
    for (int i=d_offset[u]; i<end; ++i) {
      int v = ... 
      if (d_level[v] == -1) {
        d_level[v] = *d_hops_from_source+1;
        *d_finished = false;
      }
} } }
\end{lstlisting}



In both OpenACC and SYCL, the code enters a do-while loop that executes until all the nodes have been visited. In each iteration, the code first sets a boolean variable \texttt{finished} to true, indicating that the current level has been processed. It then copies this value to the device memory and launches a kernel function in parallel to process all nodes at the current level.
The kernel function iterates through all nodes at the current level and explores their adjacent nodes by iterating over the edges that start at the current node. If an adjacent node has not been visited yet, its level is updated to the current level plus one, and the boolean variable \texttt{finished} is set to false, indicating that there are still nodes to be processed.
After processing all nodes at the current level, the code updates the level to process in the next iteration and increments the number of hops from the source node. The value of \texttt{finished} is then copied back from the device memory to the host memory to check if all the nodes have been visited.
The code continues the do-while loop until all the nodes have been visited, which is indicated by the value of the \texttt{finished} variable being \texttt{true}.


The OpenCL backend code is similar to CUDA.

\subsection{Min/Max Constructs}


SSSP in StarPlat uses a Min construct as below.

\texttt{ <nbr.dist,nbr.modified> = <Min (nbr.dist, v.dist + e.weight), True>; }   \newline
This syntax allows multiple variables (\texttt{nbr.modified and nbr.dist}) to be updated atomically.
        
Here, the for-loop that iterates through all the vertices is parallelized on the accelerator. Each iteration looks for the neighbors of one specific vertex and performs an update operation. When this is being done in parallel, there can be data-races, that is,  multiple iterations can write into the distance of the same neighbor at the same time. 

The CUDA backend handles Min/Max constructs using the atomic instructions that are readily supported in the language (that is, \texttt{atomicMin} and \texttt{atomicMax}).

To prevent the data race, OpenACC provides a  \texttt{\#pragma acc atomic write} directive. This directive ensures that no two threads / iterations write to this same location at the same time, thereby preventing data races. Figure~\ref{min-openacc} is the OpenACC generated code for the Min construct in SSSP.

\begin{lstlisting}[language=NEAR, style=mystyle, label= min-openacc,  caption = {OpenACC code generated for Min-Construct in SSSP}]
int dist_new = dist[v] + weight[e];
bool modified_new = true;
if(dist[nbr] > dist_new)
{
  int oldValue = dist[nbr];
  #pragma acc atomic write
    dist[nbr] = dist_new;

  modified_nxt[nbr] = modified_new;
  #pragma acc atomic write
      finished = false;
}
\end{lstlisting}

Figure~\ref{min-sycl} shows the SYCL code generated for \texttt{Min}. The conditional update on the node property is achieved through atomic implementations of min and max operations. 
A relaxed memory ordering is used under which the memory operations can be reordered without any restrictions.

\begin{lstlisting}[language=NEAR, style=mystyle, label= min-sycl,  caption = {SYCL code  generated for the \texttt{Min} construct}]
int nbr = gpu_edgelist[edge];
int e = edge;
int dist_new = gpu_dist[v] + gpu_weight[e];

if(gpu_dist[v]!= INT_MAX && gpu_dist[nbr] > dist_new)
{
  atomic_ref<int, memory_order::relaxed, memory_scope::device, access::address_space::global_space> atomic_data(gpu_dist[nbr]);
  atomic_data.fetch_min(dist_new);
  d_modified_next[nbr] = modified_new;
  *d_finished = false ;
}
\end{lstlisting}

\subsection{fixedPoint Construct}
The \texttt{fixedPoint} construct translates to a while loop conditioned on the fixed-point variable provided in the construct.
Typically, the convergence is based on a node property, but it can be an arbitrary computation.
This code is generated on the host, so it is similar in template for all the four backends. CUDA code updates a copy of the \texttt{finished} flag on the GPU, which is \texttt{cudaMemcpy}'ed to the host (as shown in Figure~\ref{fixedPoint-gen}).

\begin{lstlisting}[language=NEAR, style=mystyle, label= fixedPoint-gen,  caption = Code generated for the \texttt{fixedPoint} construct]
while (!finished) {
    finished = true;

    if (...) { // Min-construct expansion
        ...        
        finished = false; // fixedPoint identifier updation
}   }   
\end{lstlisting}   

\section{Backend Optimizations} \label{sec optimizations}

We now discuss the accelerator-specific optimizations.

\subsection{CUDA and OpenCL} \label{subsec opt cuda}
\mypara{Optimized Host-Device Data Transfer} 
We run a basic programme analysis on the AST to determine which variables must be transmitted between devices. For instance, since a graph is static, its copy from the GPU to the CPU at the conclusion of the kernel is not necessary. The updated vertex attributes, however, need to be returned. Similar to this, during a fixed-point processing, the \texttt{finished} flag is set on the CPU, conditionally set on the GPU, and read on the CPU once again. The variable must therefore be moved back and forth the two devices. Device-only variables are generated for the \texttt{forall}-local variables.

\mypara{Memory Optimization in OR-Reduction} 
When we write the \texttt{fixedPoint} construct, the fixed-point is computed using the \texttt{modified} attribute. If any of the vertices' \texttt{modified}  flags are set, another iteration is fundamentally required. \name makes use of this, which is essentially a logical-OR operation, to create a single flag variable that is set by multiple threads concurrently (with a reliance on hardware atomicity for primitive types). In terms of time and memory, managing this flag is less expensive than moving arrays of the \texttt{modified} flags across devices. 

\subsection{OpenACC} \label{subsec opt openacc}
\mypara{Optimized Host-Device Data Transfer} In OpenACC, data pragmas have to be generated outside each block after careful analysis of where and when data structures are used in CPU and GPU and when they have to
be transferred between CPU and GPU. The data copy between CPU and GPU is a very time-expensive operation and data transfers have to be minimised as much as possible to 
optimise program run time. Multiple consecutive GPU blocks are combined within a single data transfer so that data are copied less frquently between CPU and GPU.
StarPlat performs an analysis to find the necessary graph data variables to be copied to the accelerator before launching the kernel.

\mypara{Optimized Data Copy around Loops} For GPU parallelised loop, our analysis finds out which variables or arrays need to be copied into GPU at the start of each iteration,
which variables need to be copied out into CPU after each iteration and which variables need not be copied in or out. On the basis of this analysis, data \texttt{copyin(),
copyout(), copy()} pragmas are generated outside the loop.

\subsection{SYCL} \label{subsec opt sycl}
\mypara{Optimized Host-Device Data Transfer}
The abstract syntax tree (AST) is first analysed in the code-generation process to determine the variables that require a transfer between devices. The graph object is static, thus eliminating the need for constant copying between the GPU and CPU during kernel execution. The graph information is instead transferred to the GPU at the beginning of the function using \texttt{malloc\_device}. Properties that are modified on the GPU, such as betweenness centrality values or distance from the source for each vertex, must be sent back to the CPU after processing. Additionally, the finished flag is set on the CPU, conditionally set on the GPU, and then read again on the CPU during fixed-point processing. As a result, the variable must be transferred back and forth. Finally, the \texttt{forall}-local variables are generated as device-only variables.

\mypara{Memory Optimization in Reduction}
In the StarPlat code-generation process, the \texttt{fixedPoint} construct uses a modified property to compute the fixed-point. At a high level, another iteration is required if any of the vertices’ modified flags is set, which can be interpreted as a logical-OR operation. To optimize this process, StarPlat generates a single flag variable that is set in parallel by threads. This can be done due to the hardware atomicity for primitive types. The advantage of managing this flag variable is that it is cheaper in terms of time and memory compared to transferring arrays of modified flags across devices.

\mypara{Computing FixedPoint Efficiently}
The fixedPoint construct converges on a specific condition on a single boolean node property. The change of convergence is tracked through a boolean fixed-point variable that ideally needs to be updated after analyzing the property values for all nodes. The update procedure has been optimized by updating the fixed-point variable along with the update to the property value for any node. Since updates to a boolean variable are atomic by hardware, this does not lead to a performance loss.
\section{Experimental Evaluation} \label{sec exp evaluation}

\mypara{Algorithms} Using StarPlat, we developed four algorithms: Betweenness Centrality (BC),
PageRank (PR), Single Source Shortest Path (SSSP), and Triangle Counting (TC), for benchmarking and for comparing with other frameworks and graph libraries. The StarPlat DSL codes for BC and PR fit in 30 lines each while those for SSSP and TC fit in 20. With this small specification, a domain-expert can generate the implementations for these algorithms in different accelerator forms: CUDA, OpenACC, SYCL, OpenCL. Ignoring the header files, the compiler generates around 150, 120, 125, and 75 lines for BC, PR, SSSP, and TC respectively for the CUDA backend. The numbers reduce by about 33\% for OpenACC, and increase by 50\% and 100\% for SYCL and OpenCL. 


\mypara{Graphs} We use ten large graphs in our experiments, which are a mix of different types. Six of these are social networks exhibiting the small-world property, two are road networks having large diameters and small vertex degrees, while two are synthetically generated. One synthetic graph has a uniform random distribution (generated using Green-Marl's graph generator~\cite{greenmarl}), while the other one has a skewed degree distribution following the recursive-matrix format (generated using SNAP’s RMAT generator with parameters a = 0.57, b = 0.19, c = 0.19, d = 0.05)~\cite{snap}. They are listed in Table~\ref{graph-inputs}, sorted on the number of edges in each category. 
For unweighted graphs, we assign edge-weights selected uniformly at random in the range [1,100] (for SSSP).

\begin{table}
\centering
\small
\setlength{\tabcolsep}{4pt}
\begin{tabular}{r|c|r|r|r|r}
 \hline
 \textbf{Graph} & \textbf{Short} & \multicolumn{1}{c|}{\textbf{$|$V$|$}} &  \multicolumn{1}{c|}{\textbf{$|$E$|$}}  & \textbf{Avg. $\delta$} & \textbf{Max. $\delta$} \\
    & \textbf{name} & \multicolumn{1}{c|}{$\times 10^6$} & \multicolumn{1}{c|}{$\times 10^6$} & & \\
 \hline
 twitter-2010 & TW   & 21.2  & 265.0 & 
 12.0 & 302,779\\ 
 soc-sinaweibo & SW &   58.6  & 261.0 & 
 4.0 & 4,000 \\
 orkut & OK & 3.0 & 234.3 
 & 76.3  &33,313\\
 wikipedia-ru & WK & 3.3 & 93.3 
 & 55.4 & 283,929\\
 livejournal &LJ &  4.8 & 69.0 
 &28.3 & 22,887\\
 soc-pokec & PK & 1.6 & 30.6 
 & 37.5 & 20,518\\
 \hline
 usaroad & US & 24.0  & 28.9 
 & 2.0 & 9  \\
 germany-osm & GR & 11.5  & 12.4 
 & 2.0  & 13\\ \hline
 rmat876 & RM & 16.7 & 87.6  
 & 5.0 & 128,332\\
 uniform-random & UR & 10.0 & 80.0 
 & 8.0 & 27 \\
 \hline
\end{tabular}
\caption{Input graphs ($\delta$ indicates degree)} 

\label{graph-inputs}
\end{table}

\mypara{Machine Configuration} We ran our experiments for CUDA, OpenCL, and OpenACC on a compute node, 
whose configuration is as follows: Intel Xeon Gold 6248 CPU with 40 hardware threads spread over two sockets, 2.50~GHz clock, and 192 GB memory running RHEL 7.6 OS. 
All the codes in C++ are compiled with GCC 9.2, with optimization flag -O3. Various backends have the following versions: OpenACC version 20.7-0, OpenCL version 2.0, CUDA version 10.1.243 
and run on Nvidia Tesla V100-PCIE GPU with 5120 CUDA cores spread uniformly across 80 SMs clocked at 1.38~GHz with 32~GB global memory and 48 KB shared memory per thread-block.
Since SYCL code can run on both CPU and GPU, we ran it on the CPU (with the configuration mentioned above, with 40 threads). However, since the machine does not house Intel GPUs and is not configured to work with its V100 GPUs, we ran it on Intel DevCloud GPU UHD Graphics [0x9a60] and also on another NVIDIA GPU GeForce RTX 2080 Ti.\footnote{Note that SYCL code can run on NVIDIA GPUs by installing a plugin and satisfying certain software and hardware dependencies.}

\mypara{Baselines}
We compare StarPlat-generated accelerator codes against the following two hand-crafted baselines, which also implement the same set of algorithms (except for BC).
\begin{itemize}
    \item Gunrock~\cite{Gunrock} provides data-centric abstractions to apply a graph operator on vertices or edges to compute the next frontier using the following three functions to be supplied by the user: \textsf{filter}, \textsf{compute}, and \textsf{advance}. It supports the CUDA backend. By optimizing the processing for improved coalescing and reduced thread-divergence, the Gunrock library constructs efficient implementations of graph processing.
    \item LonestarGPU~\cite{lonestargpu} is a collection of hand-optimized CUDA programs, with varied optimizations applied for data-driven and topology-driven 
    computations~\cite{datatopology}.
\end{itemize}

\begin{table*}
\arrayrulecolor{gray}
\scalebox{0.9}{
\setlength{\tabcolsep}{3pt}
    \begin{tabular}{|lr|c|r|r|r|r|r|r|r|r|r|r||r|}
\hline
\multicolumn{2}{|l|}{Algo.} & Framework & TW & SW & OK & WK & LJ & PK & US & GR & RM & UR & Total \\ \hline
\multicolumn{1}{|l|}{\multirow{6}{*}{BC}} & 1 & LonestarGPU & - & - & - & - & - & - & - & - & - & - & - \\
\multicolumn{1}{|l|}{} & 1 & Gunrock & 2.122 & 4.237 & 0.525 & 0.535 & 0.548 & 0.317 & \textbf{2.811} & \textbf{1.750} & 1.238 & 0.944 & 15.027 \\
\multicolumn{1}{|l|}{} & 1 & \name & \textbf{0.002} & \textbf{0.004} & \textbf{0.149} & \textbf{0.153} & \textbf{0.078} & \textbf{0.029} & 17.656 & 6.359 & \textbf{0.225} & \textbf{0.079} & 24.734 \\ \cline{2-14} 
\multicolumn{1}{|l|}{} & 20 & \multirow{3}{*}{\name} & 6.992 & 2.279 & 2.762 & 3.014 & 1.298 & 0.534 & 369.701 & 126.485 & 2.949 & 1.593 & 517.607 \\
\multicolumn{1}{|l|}{} & 80 &  & 28.179 & 9.332 & 11.331 & 12.050 & 4.886 & 1.907 & 1444.656 & 518.968 & 6.509 & 6.372 & 2044.189 \\
\multicolumn{1}{|l|}{} & 150 &  & 55.548 & \MLE & 21.241 & 27.271 & 9.609 & 3.794 & 2636.453 & 978.758 & 9.912 & 11.957 & 3754.543 \\ \hline
\multicolumn{2}{|l|}{\multirow{3}{*}{PR}} & LonestarGPU & - & \textbf{0.240} & 0.363 & \textbf{0.104} & \textbf{0.225} & \textbf{0.240} & \textbf{0.832} & \textbf{0.294} & \textbf{0.240} & \textbf{0.240} & 2.778 \\
\multicolumn{2}{|l|}{} & Gunrock & 15.230 & 36.910 & 2.430 & 2.460 & 2.952 & 1.085 & 13.345 & 6.499 & 9.170 & 5.487 & 95.568 \\
\multicolumn{2}{|l|}{} & \name & \textbf{4.081} & 7.112 & \textbf{0.256} & 1.780 & 1.300 & 0.257 & 3.420 & 0.679 & 0.891 & 0.257 & 20.033 \\ \hline
\multicolumn{2}{|l|}{\multirow{3}{*}{SSSP}} & LonestarGPU & - & 0.077 & 0.217 & 0.058 & 0.084 & 0.037 & \textbf{0.162} & \textbf{0.091} & 0.129 & 0.183 & 1.083 \\
\multicolumn{2}{|l|}{} & Gunrock & 2.272 & 4.057 & 0.616 & 0.556 & 0.562 & 0.311 & 1.283 & 1.140 & 1.034 & 0.915 & 12.746 \\
\multicolumn{2}{|l|}{} & \name & \textbf{0.001} & \textbf{0.002} & \textbf{0.078} & \textbf{0.044} & \textbf{0.027} & \textbf{0.012} & 1.667 & 0.695 & \textbf{0.120} & \textbf{0.028} & 2.674 \\ \hline
\multicolumn{2}{|l|}{\multirow{3}{*}{TC}} & LonestarGPU & - & 31.990 & 2.998 & 2.771 & \textbf{0.110} & \textbf{0.039} & 11.874 & 5.695 & \textbf{1.270} & 0.499 & 57.246 \\
\multicolumn{2}{|l|}{} & Gunrock & \textbf{67.718} & 7.369 & \textbf{0.843} & \textbf{0.997} & 0.850 & 0.404 & 1.490 & 0.712 & 3.200 & 1.040 & 84.623 \\
\multicolumn{2}{|l|}{} & \name & 10540.002 & \textbf{1.410} & 46.700 & 4.009 & 3.006 & 0.655 & \textbf{0.001} & \textbf{0.001} & 824.620 & \textbf{0.034} & 11420.430 \\ \hline
\end{tabular}
}
\caption{\name's CUDA code performance comparison against LonestarGPU and Gunrock. All times are in seconds. The number in the second column for BC is the number of iterations executed. LonestarGPU does not have BC implemented and fails to load the largest graph TW ({\MLE} == out of memory)
}


\label {table:cuda}
\end{table*}

\subsection{Comparison across Frameworks}
Our goal is to match the performance of the hand-crafted code.
Table~\ref{table:cuda} presents the absolute running times of the four algorithms on our ten graphs for the three frameworks: LonestarGPU, Gunrock, and \name.
The running times include CPU-GPU data transfer (since we will be comparing these times against the CPU-only times as well).



\mypara{\bcc} BC involves running forward and backward shortest paths from various source vertices. Since Brandes algorithm~\cite{brandes} is time-consuming, literature often compares time for a few iterations, spanning a few source vertices (otherwise, it can take several days to complete BC on large graphs). The number of iterations is shown against BC in the second column in Table~\ref{table:cuda}. LonestarGPU does not have BC as part of its collection. Compared to Gunrock, \name-generated code outperforms on eight out of the ten graphs. Both the graphs for which Gunrock outperforms are road networks, having large diameter. 
Gunrock relies heavily on bulk-synchronous processing and its three API are very well optimized. Gunrock's Dijkstra's algorithm works efficiently for road networks. On the other hand, on social and random graphs, our implementation fares better. This is encouraging since StarPlat code is generated.
We also illustrate performance with multiple sources of different sizes (20, 80, and 150). Except for soc-sinaweibo, \name-generated code is on par with or better than the other frameworks. 
Finally, unlike Gunrock and LonstarGPU, \name has the provision to execute BC from a set of source vertices.

\mypara{\prr}
We observe that the three frameworks have consistent relative performance, with hand-crafted LonestarGPU codes outperforming the other two and \name outperforming Gunrock. \name exploits the double buffering approach to read the current PR values and generate those for the next iteration (see Figure~\ref{reduction-openacc}). This separation reduces synchronization requirement during the update, but necessitates a barrier across iterations.
LonestarGPU uses an in-place update of the PR values and converges faster.

\mypara{\ssspp}
Gunrock uses Dijkstra's algorithm 
for computing the shortest paths using a two-level priority queue. We have coded a variant of the Bellman-Ford algorithm in \name. Hence, the comparison may not be most appropriate. But we compare the two only from the application perspective -- computing the shortest paths from a source in the least amount of time. LonestarGPU and \name outperform Gunrock on all the ten graphs. Between LonestarGPU and \name, there is no clear winner. They, in fact, have competitive execution times.

\mypara{\tcc}
Unlike the other three algorithms, TC is not a propagation based algorithm. In addition, it is characterized by a doubly-nested loop inside the kernel (to iterative through neighbors of neighbors). Another iteration is required for checking edge which can be implemented linearly or using binary search if the neighbors are sorted in the CSR representation. Due to this variation in the innermost loop, the time difference across various implementations can be pronounced, which we observe across the three frameworks. Their performances are mixed across the ten graphs. 

\begin{table*}
\scalebox{0.9}{
\setlength{\tabcolsep}{3pt}
\begin{tabular}{|lr|c|r|r|r|r|r|r|r|r|r|r|r|}
\hline
\multicolumn{2}{|l|}{Algo.} & Framework & TW & SW & OK & WK & LJ & PK & US & GR & RM & UR & Total \\ \hline
\multicolumn{1}{|l|}{} & 1 & CUDA & \textbf{0.01} & \textbf{0.01} & \textbf{0.15} & 0.15 & 0.08 & \textbf{0.03} & 17.66 & 6.36 & \textbf{0.23} & \textbf{0.08} & 24.73 \\
\multicolumn{1}{|l|}{} & 1 & Openacc(Nvidia GPU) & {0.76} & {1.64} & {0.57} & {0.63} & {0.60} & {0.12} & 58.77 & 21.30 & {4.29} & {0.79} & 89.47 \\
\multicolumn{1}{|l|}{} & 1 & Openacc(Intel CPU) & {0.46} & {1.66} & {45.15} & {29.01} & {76.79} & {14.06} & 1670.08 & 608.26 & {103.65} & {59.13} & 2608.25 \\ 
\multicolumn{1}{|l|}{} & 1 & OpenCL(Nvidia GPU) & - & - & - & - & - & - & - & - & - & - & - \\ 
\multicolumn{1}{|l|}{} & 1 & SYCL(Intel CPU) & 0.45 & 0.51 & 1.29 & 1.08 & 0.78 & 0.58 & 71.40 & 31.09 & 2.15 & 1.25 & 110.58 \\
\multicolumn{1}{|l|}{} & 1 & SYCL(Intel GPU) & 0.21 & 0.29 & 0.86 & 0.67 & 0.35 & 0.22 & 57.37 & 21.22 & 1.11 & 1.05 & 79.35 \\
\multicolumn{1}{|l|}{} & 1 & SYCL(Nvidia GPU) & \textbf{0.01} & \textbf{0.01} & \textbf{0.15} & \textbf{0.09} & \textbf{0.07} & \textbf{0.03} & \textbf{3.02} & \textbf{1.31} & 0.56 & 0.10 & \textbf{5.34} \\ \cline{2-14} 
\multicolumn{1}{|l|}{} & 20 & {CUDA} & 6.99 & 2.28 & 2.76 & 3.01 & 1.30 & 0.53 & 369.70 & 126.49 & 2.95 & 1.59 & 517.61 \\
\multicolumn{1}{|l|}{} & 20 & Openacc(Nvidia GPU) & 16.608 & 32.6 & 11.83 & 12.3 & 6.42 & 2.3 & 1600.2 & 928.71 & 86.5 & 10.81 & 2708.278 \\
\multicolumn{1}{|l|}{} & 20 & Openacc(Intel CPU) & 8.84 & 84.5 & 983.61 & 748.49 & 1597.88 & 249.00 & 30692.97 & 13267.05 & 1778.04 & 1455.57 & 50865.95 \\
\multicolumn{1}{|l|}{} & 20 & OpenCL(Nvidia GPU) & - & - & - & - &- & - & - & -& - & - & - \\
\multicolumn{1}{|l|}{} & 20 & {SYCL(Intel CPU)} & 7.69 & 1.95 & 14.76 & 12.33 & 7.14 & 3.1 & 1611.54 & 578.56 & 32.68 & 16.35 & 2286.11 \\
\multicolumn{1}{|l|}{} & 20 & {SYCL(Intel GPU)} & 2.40 & 4.60 & 13.60 & 10.98 & 5.13 & 2.37 & 1,122.09 & 384.42 & 20.74 & 18.44 & 1584.77 \\
\multicolumn{1}{|l|}{} & 20 & {SYCL(Nvidia GPU)} & 0.10 & 0.16 & 3.06 & 1.44 & 1.11 & 0.51 & 56.69 & 23.41 & 12.22 & 1.92 & 100.60 \\ \cline{2-14} 
\multicolumn{1}{|l|}{BC} & 80 & CUDA & 28.179 & 9.332 & 11.331 & 12.050 & 4.886 & 1.907 & 1444.656 & 518.968 & 6.509 & 6.372 & 2044.189 \\ 
\multicolumn{1}{|l|}{} & 80 & Openacc(Nvidia GPU) & 63.78 & 148.28 & 0.35 & 51.91 & 371.58 & 9.23 & 7783.67 & 5644.65 & 258.06 & 36.01 & 14367.52 \\
\multicolumn{1}{|l|}{} & 80 & Openacc(Intel CPU) & 51.93 & 547.75 & 982.17 & 3166.13 & 2730.46 & 1087.26 & 96859.19 & 51268.69 & 4318.54 & 10101.74 & 171113.86 \\   
\multicolumn{1}{|l|}{} & 80 & OpenCL(Nvidia GPU) & - & - &-&- & - & - & - &-& -& -& - \\
\multicolumn{1}{|l|}{} & 80 & SYCL(Intel CPU) & 24.18 & 7.26 & 58.70 & 45.62 & 28.21 & 11.97 & 5398.65 & 2354.16 & 112.75 & 59.84 & 10455.99\\ 
\multicolumn{1}{|l|}{} & 80 & SYCL(Intel GPU) & 9.12 & 18.36 & 50.14 & 39.63 & 20.73 & 8.48 & 4196.69 & 1487.65 & 79.45 & 62.26 & 5972.51 \\ 
\multicolumn{1}{|l|}{} & 80 & SYCL(Nvidia GPU) & 0.37 & 0.68 & 11.43 & 6.07 & 4.38 & 1.87 & 206.52 & 98.62 & 47.89 & 7.37 & 385.2 \\ 
\hline

\multicolumn{2}{|c|}{}  & CUDA & 4.081 & 7.112 & \textbf{0.256} & 1.780 & 1.300 & 0.257 & 3.420 & 0.679 & 0.891 & 0.257 & 20.033 \\
\multicolumn{2}{|l|}{} & Openacc(Nvidia GPU) & \textbf{0.829} & 1.184 & \textbf{0.628} & 0.449 & 0.442 & 0.356 & 0.63 & 0.447 & 0.572 & 0.7 & 6.237 \\
\multicolumn{2}{|l|}{} & Openacc(Intel CPU) & \textbf{27.6} & 344.6 & \textbf{4.52} & 2.6 & 2.22 & 0.53 & 0.95 & 0.96 & 7.75 & 3.11 & 394.84 \\ 
\multicolumn{2}{|c|}{PR} & OpenCL(Nvidia GPU) & 345.59 & 4.49 & 33.36 & 54.12 & 55.87 & 19.77 & 204.96 & - & 439.45 & 131.87 & - \\
\multicolumn{2}{|l|}{} & SYCL(Intel CPU) & 70.22 & 39.91 & 25.92 & 13.57 & 52.95 & 21.02 & 347.02 & 197.28 & 33.96 & 47.05 & 848.90 \\
\multicolumn{2}{|l|}{} & SYCL(Intel GPU) & 14.57 & 30.72 & 4.37 & 3.28 & 7.04 & 2.76 & 43.38 & 23.53 & 7.56 & 5.69 & 142.89 \\
\multicolumn{2}{|l|}{} & SYCL(Nvidia GPU) & \textbf{2.70} & 4.00 & 1.26 & 1.19 & 0.35 & \textbf{0.15} & 1.08 & 0.55 & 1.20 & 0.30 & 12.78 \\ \hline

\multicolumn{2}{|c|}{} & CUDA & \textbf{0.001} & \textbf{0.002} & \textbf{0.078} & \textbf{0.044} & \textbf{0.027} & \textbf{0.012} & 1.667 & 0.695 & \textbf{0.120} & \textbf{0.028} & 2.674 \\
\multicolumn{2}{|l|}{} & Openacc(Nvidia GPU) & \textbf{0.72} & \textbf{0.9} & \textbf{0.65} & \textbf{0.42} & \textbf{0.36} & \textbf{0.27} & 2.8 & 1.24 & \textbf{0.96} & \textbf{0.48} & 8.8 \\ 
\multicolumn{2}{|l|}{} & Openacc(Intel CPU) & \textbf{0.995} & \textbf{----} & \textbf{0.974} & \textbf{3.1} & \textbf{0.715} & \textbf{0.452} & 3.429 & 2.22 & \textbf{3.31} & \textbf{----} & ---- \\ 
\multicolumn{2}{|c|}{SSSP} & OpenCL(Nvidia GPU) & \textbf{0.001} & \textbf{0.004} & \textbf{0.084} & \textbf{0.047} & \textbf{0.029} & \textbf{0.012} & 3.73 & - & \textbf{0.116} & \textbf{0.031} & - \\ 
\multicolumn{2}{|l|}{} & SYCL(Intel CPU) & 2.95 & 0.73 & 2.17 & 1.03 & 1.28 & 0.75 & 53.64 & 12.15 & 2.01 & 1.93 & 78.63 \\
\multicolumn{2}{|l|}{} & SYCL(Intel GPU) & 0.82 & 0.02 & 0.32 & 0.12 & 0.11 & 0.05 & 0.18 & 1.24 & 0.38 & 0.18 & 79.39 \\
\multicolumn{2}{|l|}{} & SYCL(Nvidia GPU) & 4.081 & 7.112 & 0.256 & 1.780 & 1.300 & 0.257 & 3.420 & 0.679 & 0.891 & 0.257 & 6.55 \\ \hline

 \multicolumn{2}{|l|}{} & CUDA & 10540.00 & 1.41 & 46.70 & \textbf{4.01} & 3.01 & 0.66 & \textbf{0.00} & \textbf{0.00} & 824.62 & \textbf{0.03} & \textbf{11420.43} \\ 
 \multicolumn{2}{|l|}{} & Openacc(Nvidia GPU) & 15522.84 & {3.23} & 62.00 & 6.10 & 4.98 & 1.32 & {0.40} & {0.35} & 10770.25 & {0.45} & 31625.92 \\
 \multicolumn{2}{|l|}{} & Openacc(Intel CPU) & 20776.85 & {42.10} & 3300.97 & 545.79 & 225.29 & 31.58 & {1.38} & {0.58} & 10299.78 & {15.24} & 35239.56 \\ 
\multicolumn{2}{|c|}{TC} & OpenCL(Nvidia) & \textbf{10162.17} & {1.38} & 48.04 & \textbf{4.01} & 3.08 & 0.67 & \textbf{0.00} & - & 791.77 & {0.04} & - \\ 
\multicolumn{2}{|l|}{} & SYCL(Intel CPU) & \OOT & 21.29 & 108.86 & \OOT & 11.56 & 2.21 & 0.25 & 0.19 & 1585.90 & 0.82 & -- \\
\multicolumn{2}{|l|}{} & SYCL(Intel GPU) & \OOT & 11.58 & 75.12 & \OOT & 8.76 & 2.89 & 0.13 & 0.09 & 902.38 & 0.49 & -- \\
\multicolumn{2}{|l|}{} & SYCL(Nvidia GPU) & \OOT & \textbf{0.99} & \textbf{40.81} & 4.04 & \textbf{2.37} & \textbf{0.00} & \textbf{0.00} & 0.68 & \textbf{0.89} & 0.26 & -- \\ \hline
\end{tabular}
}

\caption{StarPlat’s code performance on different accelerators. All times are in seconds. The number in the second column for BC is the number of iterations executed. ({\OOT} == one hour timeout) } 
\label{table:all}
\end{table*}

\subsection{Comparison across Accelerators} 
Technically, it is inappropriate to compare across different hardware with different software architectures. Nonetheless, we delve into this comparison from end-users' perspective who want their algorithmic code to complete execution as fast as possible.  Table~\ref{table:all} compares performance of the four algorithms across various accelerator codes generated by StarPlat. The backend CUDA is same as StarPlat from Table~\ref{table:cuda}.





\mypara{\bcc} We observe that overall SYCL on NVIDIA GPUs performs the best among the backends. On NVIDIA GPUs, OpenACC performs poorly compared to CUDA and SYCL. SYCL on NVIDIA GPU outperforms SYCL on Intel GPU. CUDA (on NVIDIA GPUs) is a close second. Unlike CUDA, SYCL's implementation does not depend upon \texttt{grid.synchronization()}, resulting in better performance on road networks. OpenACC's pragma-based implementation on NVIDIA GPU performs comparable to SYCL on Intel GPU. But on Intel CPU, OpenACC performs poorly compared to SYCL. We also observed that for short diameter graphs, the BC time scales linearly with the number of sources across the backends.

\mypara{\prr} Unlike in BC, OpenACC on NVIDIA GPUs outperforms the other backends for PR.  SYCL on NVIDIA GPUs is a close second, followed by CUDA. OpenCL on NVIDIA GPUs is the slowest (ignoring graph GR). On Intel CPU, OpenACC outperforms SYCL on all but one graph (SW). SYCL on Intel GPU outperforms the Intel CPU versions. 

\mypara{\ssspp} Similar to BC, CUDA outperforms other backends for SSSP. On NVIDIA GPUs, CUDA is followed by SYCL, then OpenACC, and then OpenCL (ignoring two graphs SW and UR for which our generated OpenCL code did not produce the correct results). On Intel CPU, OpenACC outperforms SYCL overall (similar to PR). Note that this is opposite of what we observed for BC. This suggests that algorithmic aspects play a major role in deciding the performance of hardware, and we cannot make a concluding remark about graph algorithms being most performant on a certain software-hardware pair.

\mypara{\tcc} Although small, TC is a time-consuming algorithm. This is evident from SYCL going out of time (one hour timeout) on two graphs (TW and OK). Otherwise, SYCL performs well on almost all the other graphs. On NVIDIA GPUs, CUDA and OpenCL perform similarly, while OpenACC is twice as slow. On Intel CPU, SYCL considerably outperforms OpenACC. SYCL on Intel GPU is comparable to that on Intel CPU. Among all the graphs, TW and RM stand-out for extremely high running times. This is due to (i) a large number of triangles in them, and (ii) skewed degree distribution.

\section{RELATED WORK} \label{sec relatedwork}
Graph algorithms have been mostly explored in CUDA.

Gunrock~\cite{Gunrock} is a graph library which uses data-centric abstractions to perform operations on edge and vertex frontiers. 
 All the Gunrock operations are bulk-synchronous, and they affect the frontier by operating on the values within it or by computing a new one, using the following three functions: \textit{filter}, \textit{compute}, and \textit{advance}.  Gunrock library constructs efficient implementations of frontier operations with coalesced accesses and minimal thread divergence in CUDA.
LonestarGPU~\cite{lonestargpu} is a collection of graph analytic CUDA programs. It employs multiple techniques related to computation, memory, and synchronization to improve performance of the underlying graph algorithms. We quantitatively compare \name against Gunrock and LonestarGPU.

Medusa~\cite{medusa} is a software framework which eases the work of GPU computation tasks. Similar to Gunrock, it provides APIs to build upon, to construct various graph algorithms. 
Medusa  exploits the BSP  model,  and  proposes a  new  model  EVM  (Edge  Message  Vertex), wherein  
the local  computations  are  performed  on  the  vertices and the computation progresses by passing messages across edges.  
CuSha~\cite{CUSHA} is a graph processing framework that uses two graph representations: G-Shards and Concatenated  Windows (CW). G-Shards makes use of a recently developed idea for non-GPU systems that divides a graph into ordered sets of edges called as \textit{shards}. In order to increase GPU utilisation for processing sparse graphs, CW is a new format that improves the use of shards. CuSha improves GPU utilization by processing several shards in parallel on the streaming multiprocessors. CuSha's architecture for parallel processing of large graphs allows the user to create the vertex-centric computation and plug it in, making programming easier.
CuSha is demonstrated to significantly outperform the virtual warp-centric approach.
MapGraph~\cite{MapGraph} is a parallel graph programming framework which provides a high level abstraction, which helps in writing efficient graph programs. It uses SOA (Structure Of Arrays) to ensure coalesced memory accesses. It uses the dynamic scheduling strategy using GAS (Gather-Apply-Scatter) abstraction. Dynamic scheduling improves the memory performance and dispense the workload to the threads in accordance with degree of vertices.


T. Hoshino et al.~\cite{Hoshino} presents an early comparison of OpenACC and CUDA performance using two small bencharks: stencil and matrix multiplication. They also present a real-world CFD application benchmark for comparison.
Sandra Wienke et al.~\cite{Sandra} showed a study of OpenACC and compared its performance against OpenCL on two real world applications: Simulation of bevel gear cutting and Neuromagnetic Inverse Problem. They conclude that OpenACC offers a promising development effort to performance ratio based on these benchmarks.
K. Alsubhi et al.~\cite{Alsubhi}  proposed a tool to translate sequential C++ code into OpenACC parallelised code. The tool used an analyser that detects blocks that could be run in parallel, finds data dependency between different blocks and finds the type of parallelism that could be used.


Tomusk~\cite{openclgraph} makes a case for using OpenCL for graph algorithms, and claims that OpenCL is expressive enough to support custom optimizations for graph algorithms. 
\section{CONCLUSION} \label{sec conclusion}
We illustrated that it is feasible to generate efficient parallel code for multiple accelerators from the same algorithmic specification in a graph DSL. This is not only viable, but is also a desirable approach especially for domain-experts who are currently forced to learn multiple languages for accelerating their scientific computation. A limitation of StarPlat is that it is still a new language. While writing code in the DSL is relatively easy and short, it would be helpful if the learning curve can be further reduced.

\begin{acks}
We gratefully acknowledge the use of the computing resources at HPCE, IIT Madras. This work is supported by grants from KLA and India's National Supercomputing Mission.
\end{acks}

\bibliographystyle{ACM-Reference-Format}
\bibliography{9REF}

\end{document}